\documentclass{article}
\usepackage{graphicx} 

\usepackage{PRIMEarxiv}

\usepackage{algorithm}
\usepackage{appendix}
\usepackage{algpseudocode}
\usepackage{amsmath}
\usepackage{amssymb}
\usepackage{graphicx}
\usepackage{subcaption}
\usepackage{booktabs}
\usepackage{tabularx}
\usepackage{xcolor}
\usepackage{tikz}
\usepackage{pgfplots}
\pgfplotsset{compat=1.18}
\usepackage{amsthm}

\algrenewcommand\algorithmicrequire{\textbf{Input:}}
\algrenewcommand\algorithmicensure{\textbf{Output:}}

\newcommand{\bs}[1]{\boldsymbol{#1}}

\usepgfplotslibrary{statistics}

\title{Data-free neural PDE solvers based on Graph Neural Networks and weak forms}
\author{Mikel M. Iparraguirre, Ic\'iar Alfaro, David Gonz\'alez, El\'ias Cueto\vspace{0.5cm}\\  \small Keysight-UZ Chair of the Spanish National Strategy on AI\\ \small Aragon Institute of Engineering Research \\ \small Universidad de Zaragoza, Spain \vspace{0.5cm} \\
  \texttt{\{mikel.martinez,iciar,gonzal,ecueto\}@unizar.es}}

\begin{document}

\maketitle

\begin{abstract}
{We present a physics-informed, data-free neural solver for many-query problems in quasi-static, finite-strain hyperelasticity in three dimensions. 
A Graph Neural Network with physics-attention predicts the nodal displacements on unstructured tetrahedral meshes and is trained without simulation data by minimising the Galerkin residual of the weak form, complemented with boundary-condition and element-inversion penalty terms. Spatial derivatives are obtained from finite-element shape-function gradients rather than by automatic differentiation with respect to the coordinates. The trained model generalises to unseen load cases and geometries, reaching maximum nodal force residuals below $1\%$ of the mean internal nodal force, the equilibrium criterion of standard non-linear finite element solvers.

Since this residual can be evaluated for any prediction at negligible cost, it is used at inference as an equilibrium check: when the user-defined tolerance is not met, the network is fine-tuned on the new query with the same physics-based loss.

The approach thus avoids the costly process of generating, curating and storing high-fidelity simulation data. While data-free training is not new, to the best of our knowledge this is the first time a Galerkin residual loss with an element-inversion penalty has been combined with a mesh-based geometric learning architecture for many-query problems in three-dimensional, finite-strain solid mechanics on unstructured meshes.\par}

\end{abstract}

\section{Introduction}

\subsection{{Motivation}}

The emergence of data-driven methods in computational mechanics has brought about a revolution in the way we view numerical methods and their requirements. In just a few years---less than a decade---there has been spectacular progress in the development of methodologies for practically all the governing equations of known physics and, in particular, of engineering \cite{montans2023machine}.

Traditionally, these methods are trained using synthetic data derived from high-fidelity simulations, which are in turn carried out using modern finite element, finite difference or finite volume software. However, now that the aforementioned methods of scientific machine learning have advanced rapidly and demonstrated their robustness (although they have yet to reach full maturity) and, above all, their tremendous speed of inference, the industry has begun to realise that the data required for training is, in fact, expensive to obtain, curate and store. This is why interest has begun to grow across various sectors in methods that do not require data for training. In itself, this possibility is not new, as the method largely responsible for our community’s enthusiasm for machine learning—physics-informed neural networks—does not necessarily require data (though it can be incorporated if desired) for training \cite{RAISSI2019}.

Finite Element Methods, FEM, when applied to non-linear problems, typically rely on Newton-Raphson (NR) or equivalent iterative solvers, where the objective is to find the {vector of nodal displacements} $\bs{u}$ that satisfies the equilibrium (force balance) equation
$$\bs{R}(\bs{u})=\bs{f}_{\mathrm{ext}}-\bs{f}_{\mathrm{int}}(\bs{u})=\bs{0},$$
together with the applied Dirichlet boundary conditions.
Since the internal force vector $\bs{f}_{\mathrm{int}}(\bs{u})$ is in general a non-linear function of $\bs{u}$, the equation cannot be solved directly. Instead, we usually linearise the residual $\bs{R}(\bs{u})$ around a known state $\bs{u}_k$ using a first-order Taylor expansion,
$$\bs{R}(\bs{u}_{k} + \delta \bs{u}) \approx \bs{R}(\bs{u}_k) + \left. \frac{\partial \bs{R}}{\partial \bs{u}} \right|_{\bs{u}_k} \delta \bs{u} = \bs{0}.$$
The gradient of the residual is given by the tangent stiffness matrix (i.e., $\bs{K}_T = -\frac{\partial \bs{R}}{\partial \bs{u}}$), which is derived from kinematics and the constitutive model. Implicit approaches to non-linear problems require iterative strategies: each load step must be divided into smaller steps, and within each step, multiple NR iterations are required to achieve convergence. These iterations become a major bottleneck in computational mechanics simulations. It is not only the resolution of a non-linear problem that involves multiple iterations to arrive at the solution. Often, problems such as those arising from optimisation, inverse problems, etc., require the solution of models that are only slightly different from one another but which, when repeated hundreds or thousands of times, end up incurring a tremendous computational cost. Traditionally, it was said that the industry was prepared to tolerate problems whose solution would take a night (from the moment an engineer runs a calculation at the end of the working day until they receive the solution upon arriving at work the following morning). This is no longer acceptable in an industry that is increasingly driven by market demands to develop competitive products in a fraction of the time previously required to do so.

In this context, data-driven methods have emerged as an appealing alternative for many-query problems. By learning from simulation data, machine learning models effectively map input parameters (such as geometry, boundary conditions, or material properties) directly to the corresponding solution fields. Once trained, the model can predict unseen queries in a single forward pass based on the learned priors. This represents a paradigm shift that eliminates the costly iterations of high-fidelity solvers. These powerful regression capabilities are due to the non-linearity of deep learning models, which can perform regressions with spatial and temporal resolutions for which the finite element method (FEM) would currently take an unacceptably long time. Nevertheless, despite their fast inference, pure data-driven models suffer from two critical limitations. First, the costs of acquiring offline data, which consist of performing a large number of finite element simulations to construct the training dataset. Second, from a methodological perspective, purely data-driven models lack the rigorous physical constraints that ensure physical laws are satisfied during inference. Complying with certain fundamental laws of physics, such as Galilean invariance, requires the use of techniques such as data augmentation, which ultimately comes at a prohibitive cost.

\subsection{{State of the art}}

Most scientific machine learning approaches can be categorised into three pillars: (i) the target strategy, which defines the 
input–output relationship; (ii) the spatial representation, which determines the model architecture; and 
(iii) the supervision paradigm, which defines the loss function.

In the target strategy, the current literature on solid mechanics simulation is dominated by \textit{autoregressive} 
approaches that focus on discrete, sequential time evolution, in which the model acts as a temporal integrator 
\cite{sanchezgonzalez2020}. However, Neural Operators (e.g., FNO, GNO, DeepONet) \cite{li2020fourier, LULU2021, hao2023gnot, tiwari2025LaMO, wang2021} 
focus on \textit{continuous-domain} regression,  without requiring, for instance, integration over time, thereby avoiding the 
typical error accumulation of autoregressive models by directly mapping the solution to a given target physical state $t$. 
In this work, we adopt {this} last strategy.

In the spatial representation, various works have developed architectures that introduce geometric biases by exploiting 
mesh representations of simulation data, such as Graph Neural Networks (GNNs) \cite{battaglia2018, sanchezgonzalez2020}.  
One of the major advantages of mesh-based architectures is their ability to handle varying geometries and 
boundary conditions, as interactions propagate through the mesh via a message-passing algorithm. 
In solid mechanics, MeshGraphNet (MGN) \cite{pfaff2021} stands out allowing to process node, edge, and contact attributes 
that encode physical information through learnable Message-Passing Neural Networks (MPNNs). 
However, standard MPNNs are susceptible to the under-reaching problem, struggling to efficiently propagate 
long-range spatial interactions across large meshes within a number of message-passing iterations \cite{tesan2026}.
While multi-scale and hierarchical variants have been introduced to mitigate this bottleneck, 
scaling to large computational domains remains a significant challenge \cite{zhao2026recurrent,lei2025m4gn,cao2026,han2022}. 
It is true that there are earlier architectures that utilise both a graph network and the weak form of the problem, however, 
the aforementioned limitations of the message-passing mechanism prevent the examples 
considered from accommodating complex meshes, which do not exceed a few dozen elements in 3D \cite{gao2022physics}.

As an alternative, Transformer architectures have emerged as a scalable alternative to mesh-based models. 
This point-cloud approach is capable of handling large industrial-scale meshes by capturing long-range 
node interactions. For instance, the Transolver\cite{wu2024Transolver} and Transolver++ \cite{luo2025transolver++} 
architectures combine a spatial slicing technique with an efficient physics-attention mechanism to handle varying
geometries scaling to large meshes \cite{adams2026geotransolver, alkin2025ABUPT}. 
However, the limitations of architectures based on the use of Transformers for learning about dynamic 
phenomena are also well known \cite{duthe2025mechanistic}.

Blending strengths from both worlds, MeshGraphNet-Transformer (MGN-T)\cite{iparraguirre2026MGNT} combines MPNN blocks as local geometric processors 
and a physics-attention transformer as a global processor, enabling mesh-based architectures to scale effectively 
to industrial-sized problems. This is the architecture chosen in this work.

From a supervision perspective, purely data-driven models offer no guarantee that physical laws 
will be satisfied during inference, as they lack rigorous physical constraints. 
To improve physical consistency, several works have incorporated soft physical constraints 
into the training loss in different flavours, such as thermodynamic constraints, conservation l
aws or constitutive law restrictions \cite{greydanus2019,cranmer2020,hernandez2021,cueto2023thermodynamics,bermejo2025, Elsayed2026}. 
Also, structure-preserving models enforce symmetries and energy conservation as 
hard constraints \cite{hernandez2021,hesthaven2021, David2023}. Despite the benefits of 
incorporating these inductive biases, the methods still rely on data and do not guarantee exact 
physics satisfaction during inference.

To address the data-generation bottleneck, Physics-Informed Neural Networks (PINNs) 
integrate governing differential equations directly into the neural network's loss function \cite{RAISSI2019}. 
This supervision strategy relies on the satisfaction of the PDE in strong form, calculated via automatic differentiation, evaluated at collocation points. While pioneering this strong-form paradigm, it introduces severe optimisation challenges, often leading to training pathologies due to unbalanced backpropagated gradients between the PDE residual and boundary-condition losses \cite{arora2022, rathore2024pinnsloss}, coupled with the inherent ill-conditioning of the underlying differential operators \cite{rathore2024pinnsloss}. Beyond optimisation, PINNs also face challenges in solving solid mechanics problems because their Euclidean solution space struggles to represent complex geometries. In other words, they lack {a} proper geometric bias \cite{li2025finitePINNs}.

{Data-free approaches, \textit{i.e.} methods that do not require training data, 
differ along three axes that largely determine their performance in solid mechanics.
The first is the \emph{form of the physics loss}: strong-form collocation (PINNs \cite{RAISSI2019}, 
physics-informed DeepONets \cite{wang2021} and neural operators \cite{li2024pino}),
minimisation of the total potential energy (Deep Ritz and Deep Energy Methods \cite{e2017deepritzmethoddeep, samaniego2020energy, nguyen2020deep}),
or a Galerkin weak-form residual \cite{kharazmi2019vpinn, badia2024feinn, gao2022physics, Wurth2024, wen2026, zhu2026WINO}.
The second is \emph{how spatial derivatives are obtained}: by automatic differentiation of the network output,
or by finite-element shape-function gradients on a mesh.
The third is the \emph{spatial representation}: a coordinate-based network that solves a single instance, 
a neural operator defined on a
regular grid, or a graph defined on the body-fitted mesh itself.\par}

{Variational approaches avoid high-order automatic differentiation by working with an integral 
form of the equations, obtained by integration by parts.
The Deep Ritz \cite{e2017deepritzmethoddeep} and Deep Energy Methods \cite{samaniego2020energy, nguyen2020deep} 
minimise the total potential energy,
which for hyperelastic solids provides a natural loss at finite strains. 
Since the minimum value of $\Pi$ is unknown a priori, however, the loss gives
no direct measure of how far a prediction is from equilibrium. 
Variational PINNs (VPINNs) \cite{kharazmi2019vpinn} instead minimise a Petrov--Galerkin residual,
in which the network is the trial function and a separate set of functions acts as test functions. 
Interpolated variants project the network output onto a conforming finite element space, so that derivatives and 
integrals of the trial function are evaluated exactly with the shape functions, at the cost of introducing a 
mesh \cite{badia2024feinn}. A well-known limitation of this family is that the Euclidean norm of the residual 
vector is not equivalent to the error: equivalence requires measuring the residual in the dual norm
of the test space, as done in Robust VPINNs \cite{rojas2024robust}, and the parametric setting raises further difficulties \cite{bachmayr2024variationally}.
In all these methods, the network is trained for one boundary-value problem at a time.
Our loss belongs to the same family: the network predicts nodal displacements, which are interpolated with 
linear finite element shape functions that also act as test functions
(Galerkin formulation), so that the residual loss term is the squared Euclidean norm of the standard finite element nodal residual, as in interpolated VPINNs \cite{badia2024feinn}.
It therefore shares their advantages (exact derivatives and quadrature, no automatic differentiation with respect to the coordinates) and their limitations (no equivalence with the error
and the poor conditioning of residual losses). 
Where our method differs is in the addition of the negative Jacobian penaliser to avoid inverted elements 
and the neural network of choice, which allows to solve many-query 3D problems to generalise to unseen loading 
cases and geometries with a single network, 
rather than one network per boundary-value problem.\par}

{Several recent works bring the weak form to architectures that learn families of problems. 
Besides PI-GGN \cite{gao2022physics}, discussed above, PI-MGN \cite{Wurth2024} trains a MeshGraphNet on the implicit 
finite-element residual of transient, non-linear heat transfer and generalises to unseen part geometries. 
TensorGalerkin \cite{wen2026} provides a fast tensorised Galerkin assembly and uses it to train an autoregressive graph network 
for two-dimensional wave and Allen--Cahn problems on a fixed mesh. WINO \cite{zhu2026WINO} is the closest work to ours: it trains a 
Fourier neural operator (FNO) with the squared weak-form residual of a compressible neo-Hookean solid, and uses its predictions to 
warm-start Newton iterations. Its examples are limited to two-dimensional plane-strain problems on a Cartesian background grid, 
whose cost grows rapidly with resolution in three dimensions, and its authors report an ill-conditioned
optimisation landscape for mixed boundary conditions. Graph-based weak-form solvers \cite{gao2022physics, Wurth2024}
operate directly on unstructured meshes, but have so far addressed scalar or linear problems (heat transfer, wave propagation, small-strain elasticity)
on meshes of modest size.\par}

\subsection{{Contributions}}

{Our approach combines the regression capabilities of deep learning with the equilibrium equations of conventional 
finite element methods.} The network is trained without simulation data by minimising a physics-based loss derived from 
the residual of the governing partial differential equation. During training, our method simultaneously solves a family 
of PDE-constrained solid mechanics problems, and finds the best-fitting network parameters. Once trained, it enables near-instant 
inference on unseen queries in a single forward pass. Our contributions are as follows:
\begin{itemize}
    \item {A \textbf{physics-informed loss for solid mechanics}: the nodal residual computed from the predicted nodal displacements and weak formulation, complemented with boundary-condition and negative-Jacobian penalties that prevent element inversion under large deformations.}
    \item {A \textbf{mesh-based, many-query neural solver}. Spatial derivatives are obtained from finite-element shape-function gradients rather than by automatic differentiation with respect to the coordinates, and the MeshGraphNet-Transformer architecture \cite{iparraguirre2026MGNT} handles three-dimensional unstructured meshes with varying geometries and boundary conditions.}
    \item {\textbf{Test-time fine-tuning driven by an equilibrium check}. The nodal force residual of any prediction is available at negligible cost; if it exceeds a user-defined tolerance at inference, the network is fine-tuned on that query with the same physics-based loss until the tolerance is met. This certifies discrete equilibrium to the prescribed tolerance.}
\end{itemize}

{In this work, we take three-dimensional large-deformation hyperelasticity, with a neo-Hookean material, as the model problem. Our method differs from previous ones in three respects: (i) it operates on unstructured, body-fitted tetrahedral meshes, thanks to the geometric bias and global attention of MGN-T; (ii) its loss is built on nodal force residuals rather than on the potential energy, so its target value is known (zero) and convergence can be checked with the same equilibrium criterion used by finite element codes; and (iii) this check is used to fine-tune the network itself at test time, instead of handing the prediction over to a classical solver.\par}

\subsection{{Scope and limitations}}

{The class of problems addressed is deliberately restricted to quasi-static, finite-strain hyperelasticity of homogeneous, compressible neo-Hookean solids in three dimensions, loaded through prescribed displacements and discretised with linear tetrahedra with a single integration point. The formulation itself is not tied to these choices; higher-order elements, other constitutive laws, including path-dependent ones, and heterogeneous materials are left for future work (Section~\ref{sec:future}). As for any residual-based criterion, a small residual certifies discrete equilibrium rather than bounding the displacement error. Note, however, that the loss is not minimised directly over $\bs{u}$ for a single problem, but through the parameters of a network trained over a family of queries.\par}

{The structure of the paper is as follows. In Section \ref{sec:background} we introduce the weak form of the problem, its finite element discretisation and the equilibrium check used throughout the paper. In Section \ref{sec:meth} we present the basic ingredients of the proposed methodology. The examples considered are described in detail in Section \ref{sec:examples}. In Section \ref{sec:experiments} we provide the results that demonstrate the performance of the method. Section~\ref{sec:future} outlines future work, and Section~\ref{sec:conclusion} concludes the paper.\par}

\section{Background}
\label{sec:background}

\subsection{{Continuous problem and weak form}}

{Let $\Omega_0\subset\mathbb{R}^3$ denote the reference configuration of the body, with boundary $\partial\Omega_0=\Gamma_D\cup\Gamma_N$, $\Gamma_D\cap\Gamma_N=\emptyset$. In a total Lagrangian description \cite{belytschko2014nonlinear}, quasi-static equilibrium reads
\begin{equation}
\label{eq:strong}
-\nabla_0\cdot\bs{P}(\bs{F}(\bs{u})) = \bs{f} \ \text{in } \Omega_0, \qquad
\bs{u}=\bar{\bs{u}} \ \text{on } \Gamma_D, \qquad
\bs{P}\,\bs{N}_0=\bar{\bs{t}} \ \text{on } \Gamma_N,
\end{equation}
where $\bs{u}$ is the displacement field, $\bs{F}=\bs{I}+\nabla_0\bs{u}$ the deformation gradient, $\bs{P}=\partial\Psi/\partial\bs{F}$ the first Piola--Kirchhoff stress derived from a stored energy density $\Psi$, $\bs{N}_0$ the outward unit normal in the reference configurtion, and $\bs{f}$, $\bar{\bs{u}}$ and $\bar{\bs{t}}$ the body force, prescribed displacement and prescribed traction, respectively.

We introduce the trial and test spaces
\begin{equation*}
\mathcal{U}=\{\bs{v}\in[H^1(\Omega_0)]^3:\ \bs{v}=\bar{\bs{u}} \text{ on } \Gamma_D\}, \qquad
\mathcal{V}=\{\bs{v}\in[H^1(\Omega_0)]^3:\ \bs{v}=\bs{0} \text{ on } \Gamma_D\}.
\end{equation*}
Multiplying (\ref{eq:strong}) by a test function $\bs{v}\in\mathcal{V}$, integrating over $\Omega_0$ and integrating by parts gives the weak form, as the boundary term on $\Gamma_D$ vanishes because $\bs{v}=\bs{0}$ there. Find $\bs{u}\in\mathcal{U}$ such that
\begin{equation}
\label{eq:weak}
\int_{\Omega_0}\bs{P}(\bs{F}(\bs{u})):\nabla_0\bs{v}\,d\Omega
-\int_{\Omega_0}\bs{f}\cdot\bs{v}\,d\Omega
-\int_{\Gamma_N}\bar{\bs{t}}\cdot\bs{v}\,d\Gamma = 0
\qquad \forall\,\bs{v}\in\mathcal{V}.
\end{equation}
Integration by parts transfers one derivative to the test function, so that only first derivatives of $\bs{u}$ are required. For hyperelastic materials, (\ref{eq:weak}) is the stationarity condition of the total potential energy \cite{zienkiewicz2005}
\begin{equation}
\label{eq:energy}
\Pi(\bs{u})=\int_{\Omega_0}\Psi(\bs{F}(\bs{u}))\,d\Omega-\int_{\Omega_0}\bs{f}\cdot\bs{u}\,d\Omega-\int_{\Gamma_N}\bar{\bs{t}}\cdot\bs{u}\,d\Gamma,
\end{equation}
whose minimiser over $\mathcal{U}$ is the physical displacement field.\par}

\subsection{{Finite element discretisation}}

{The reference domain is discretised into a mesh of $n_{\mathrm{el}}$ linear tetrahedra $\Omega_e$ with $n_{\mathrm{nod}}$ nodes $\bs{X}_i$. With each node we associate a global basis function $\varphi_i$, $i=1,\dots,n_{\mathrm{nod}}$, which is continuous, piecewise linear and satisfies $\varphi_i(\bs{X}_j)=\delta_{ij}$. Continuity across elements makes the discrete spaces conforming ($\mathcal{V}^h\subset\mathcal{V}$), so that the integration by parts leading to (\ref{eq:weak}) introduces no inter-element jump terms. The discrete trial and test spaces are
\begin{equation*}
\mathcal{U}^h=\Big\{\bs{u}^h=\sum_{i=1}^{n_{\mathrm{nod}}}\varphi_i\,\bs{u}_i:\ \bs{u}_i=\bar{\bs{u}}(\bs{X}_i)\ \text{for } \bs{X}_i\in\Gamma_D\Big\},\qquad
\mathcal{V}^h=\Big\{\bs{v}^h=\sum_{i=1}^{n_{\mathrm{nod}}}\varphi_i\,\bs{v}_i:\ \bs{v}_i=\bs{0}\ \text{for } \bs{X}_i\in\Gamma_D\Big\},
\end{equation*}
where, for nodes on which only some components are prescribed, the conditions apply to those components only. Each $\bs{u}^h\in\mathcal{U}^h$ is identified with its vector of nodal displacements $\bs{u}=[\bs{u}_1;\dots;\bs{u}_{n_{\mathrm{nod}}}]\in\mathbb{R}^{3n_{\mathrm{nod}}}$. The deformation gradient of the discrete field is
\begin{equation}
\label{eq:Fh}
\bs{F}^h:=\bs{F}(\bs{u}^h)=\bs{I}+\sum_{i=1}^{n_{\mathrm{nod}}}\bs{u}_i\otimes\nabla_0\varphi_i .
\end{equation}

Choosing as test functions the same basis functions, $\bs{v}^h=\varphi_i\,\bs{e}_k$, the weak form (\ref{eq:weak}) yields one vector equation per node,
\begin{equation}
\label{eq:residual}
\bs{R}_i(\bs{u}) := \bs{f}_{\mathrm{ext},i}-\bs{f}_{\mathrm{int},i}(\bs{u}) = \bs{0}, \qquad
\bs{f}_{\mathrm{int},i}=\int_{\Omega_0}\bs{P}(\bs{F}^h)\,\nabla_0\varphi_i\,d\Omega, \quad
\bs{f}_{\mathrm{ext},i}=\int_{\Omega_0}\bs{f}\,\varphi_i\,d\Omega+\int_{\Gamma_N}\bar{\bs{t}}\,\varphi_i\,d\Gamma,
\end{equation}
which must hold for all components that are not prescribed. We denote by $\bs{u}^\star$ the solution of this discrete problem, which equivalently minimises $\Pi$ over $\mathcal{U}^h$.\par}

\subsection{{Element-level computation}}

{Restricted to an element $\Omega_e$, the only non-vanishing basis functions are the four element shape functions $N^e_a=\varphi_{i(e,a)}|_{\Omega_e}$, $a=1,\dots,4$, where $i(e,a)$ maps local to global node numbers. For linear tetrahedra, $\nabla_0 N^e_a$ is constant in each element, and so are $\bs{F}^h$ and $\bs{P}$; one-point quadrature is therefore exact and the elemental internal forces are
\begin{equation}
\label{eq:force_integral}
\bs{f}^{e}_{\mathrm{int},a}=\int_{\Omega_e}\bs{P}\,\nabla_0 N^e_a\,d\Omega = V^e\,\bs{P}^e\,\nabla_0 N^e_a,
\end{equation}
where $V^e$ is the volume of the element and $\bs{P}^e$ its (constant) stress. The nodal internal forces follow by assembly over the set $\mathcal{E}(i)$ of elements sharing node $i$,
\begin{equation}
\label{eq:residual_aasembly}
\bs{f}_{\mathrm{int},i}=\sum_{e\in\mathcal{E}(i)}\bs{f}^{e}_{\mathrm{int},a(e,i)},
\end{equation}
with $a(e,i)$ the local number of node $i$ in element $e$. The gradients $\nabla_0 N^e_a$ and volumes $V^e$ depend only on the reference mesh and are computed once. In all the examples of this paper $\bs{f}=\bs{0}$ and $\bar{\bs{t}}=\bs{0}$, and the loading enters only through $\bar{\bs{u}}$; hence $\bs{R}_i=-\bs{f}_{\mathrm{int},i}$ at the free degrees of freedom, whereas at the prescribed ones $\bs{f}_{\mathrm{int},i}$ are the reaction forces.\par}

\subsection{Application to compressible Neo-Hookean materials}
\label{sec:neo_hook}

All the examples to be presented involve a neo-Hookean material. {We consider it a representative hyperelastic model at finite strains; the formulation applies to any hyperelastic law through $\bs{P}=\partial\Psi/\partial\bs{F}$, but other material classes have not been tested.} For completeness, we describe its compressible version in detail. Neo-Hookean material models describe a particular nonlinear stress–strain response of hyperelastic materials undergoing large deformations. In our case, the strain energy density function is assumed to be of the type
\begin{equation*}
\Psi =
\frac{\mu}{2}\left(\mathrm{tr}(\bs{C}) - 3\right)
- \mu \ln J
+ \frac{\lambda}{2} (\ln J)^2,
\end{equation*}
where $\lambda$ and $\mu$ are the Lamé constants, $\bs{C}$ is the right Cauchy-Green tensor, $\bs{C} = \bs{F}^T\bs{F}$ and $J = \text{det}(\bs{F})$ the Jacobian of the transformation, measuring volume change.

The strain energy potential can be differentiated so as to provide its conjugate stress measure, leading to the second Piola-Kirchhoff tensor (PK2),
\begin{equation*}
\bs{S}
= 2
\frac{\partial \Psi(\bs{C})}{\partial \bs{C}}
=
\mu \left( \bs{I} - \bs{C}^{-1} \right)
+
\lambda (\ln J)\,\bs{C}^{-1}.
\end{equation*}
Alternate stress measures can be derived, such as the  first Piola-Kirchhoff tensor $\bs{P}$,  often referred to as PK1, and the Cauchy stress tensor $\bs{\sigma}$,
\begin{equation*}
\bs{P}
=
\bs{F} \bs{S}; \quad \bs{\sigma}=J^{-1}\bs{P} \bs{F}^T.
\end{equation*}

\subsection{{Equilibrium check}}
\label{sec:residual_criterion}

{The residual (\ref{eq:residual}) can be evaluated for any vector of predicted nodal displacements $\hat{\bs{u}}$. 
As in standard non-linear finite element solvers, where a force-equilibrium check of this type is used as convergence criterion, a prediction is accepted when
\begin{equation}\label{sec:appendix_criteria}
r_{\max}=\frac{R_{\max}}{\tilde{q}}\le 0.01, \qquad R_{\max}=\max_{(i,k)\in\mathcal{D}_{\mathrm{free}}}|R_{ik}(\hat{\bs{u}})| ,
\end{equation}
where $R_{ik}$ is the $k$-th component of the residual at node $i$ and $\mathcal{D}_{\mathrm{free}}$ is the set of free degrees of freedom, 
so that $R_{\max}$ is the largest residual component. Constrained degrees of freedom are excluded from $R_{\max}$, since their internal forces 
are reactions rather than residuals, whereas the free components of partially constrained nodes are retained. The reference force magnitude $\tilde{q}$ 
is defined as the mean of the internal force component magnitudes $|f_{\mathrm{int},ik}|$ over the active components, \textit{i.e.} 
those satisfying $|f_{\mathrm{int},ik}|>\epsilon\max_{j,l}|f_{\mathrm{int},jl}|$, with $\epsilon=10^{-5}$. The average includes the constrained degrees 
of freedom, where $\bs{f}_{\mathrm{int},i}$ are the reaction forces; since these remain finite at equilibrium while the residuals at the free nodes vanish, 
$\tilde{q}$ represents the level of the load-carrying forces. The ratio $r_{\max}$ is dimensionless because it compares two forces; it does not divide the 
residual by the displacement field, and the tolerance of $1\%$ quoted in this paper refers to this ratio. The threshold $\epsilon$ only discards components 
whose internal force is numerically zero, which would otherwise dilute the average; for a value this small, $\tilde{q}$ is insensitive to its precise choice.\par}

\section{Methodology}\label{sec:meth}

The proposed framework includes: (i) a neural operator {that approximates the parameter-to-solution map of the problem, \textit{i.e.} the map from a query (geometry and prescribed displacements) to the nodal displacements}; (ii) a spatial representation based on a hybrid MGN–T architecture; and (iii) a supervision paradigm, defined by a physics-informed, data-free loss function.

\subsection{Target strategy: continuous regression}

\begin{figure}[h!]
    \centering
    \includegraphics[width=\linewidth]{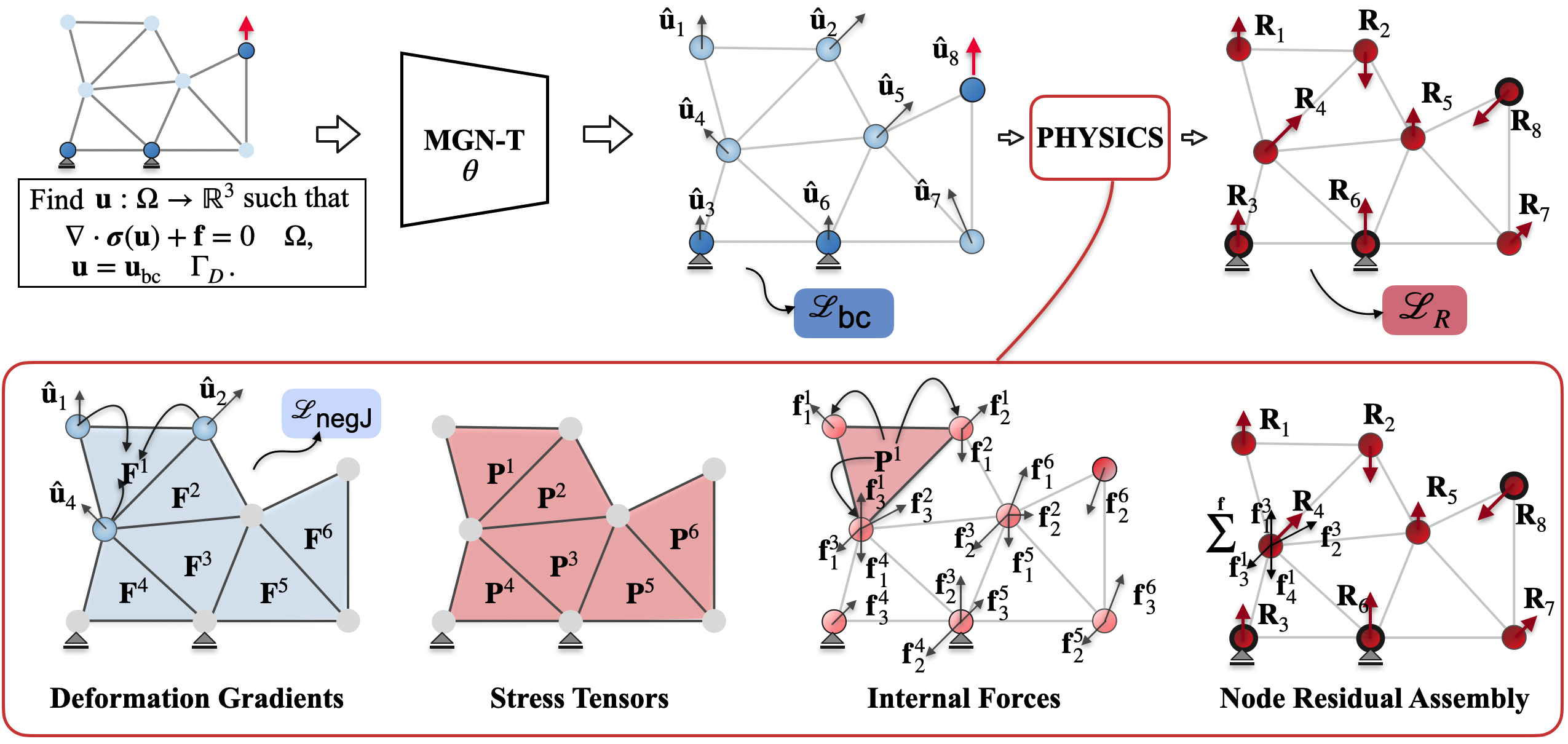}
    \caption{Sketch of the proposed method. Given a finite element mesh (top row, left), this is modelled as a graph neural network structure which, during training, will propose the set of nodal displacement vectors. From these, the loss functional is calculated, comprising one term associated with the boundary conditions and another associated with the residual at each node. To calculate the latter (the box in the bottom row), the various strain measures required will be calculated from the nodal displacements using the finite element method, and their conjugate stress measures will be determined using the constitutive equations. This will enable the residual to be determined, which is the quantity to be minimised by backpropagation.}
    \label{fig:pipeline_physics_calculation}
\end{figure}

{A query $q=(\mathcal{T}_h,\bar{\bs{u}})$ consists of a mesh and the corresponding prescribed displacements. For any query, the model predicts the vector of nodal displacements in a single forward pass,
\begin{equation}
\label{eq:prediction}
\hat{\bs{u}}_{\bs{\theta}}(q)=\text{MGN-T}\big(G(q),\bar{\bs{u}};\bs{\theta}\big),
\end{equation}
where $G(q)=(V,E)$ is the graph induced by the mesh, with nodes $V$ and edges $E$, and $\bs{\theta}$ are the network parameters (see Fig. \ref{fig:pipeline_physics_calculation}). The dependence on $\bs{\theta}$ is written as a subscript throughout and omitted only when it is clear from the context.\par}

To ensure consistency, strain and stress tensors are directly computed from the predicted displacement using the equations described in Section \ref{sec:background}.

\subsection{Spatial Representation: MGN-T}

\begin{figure}[h!]
    \centering
        \includegraphics[width=\linewidth]{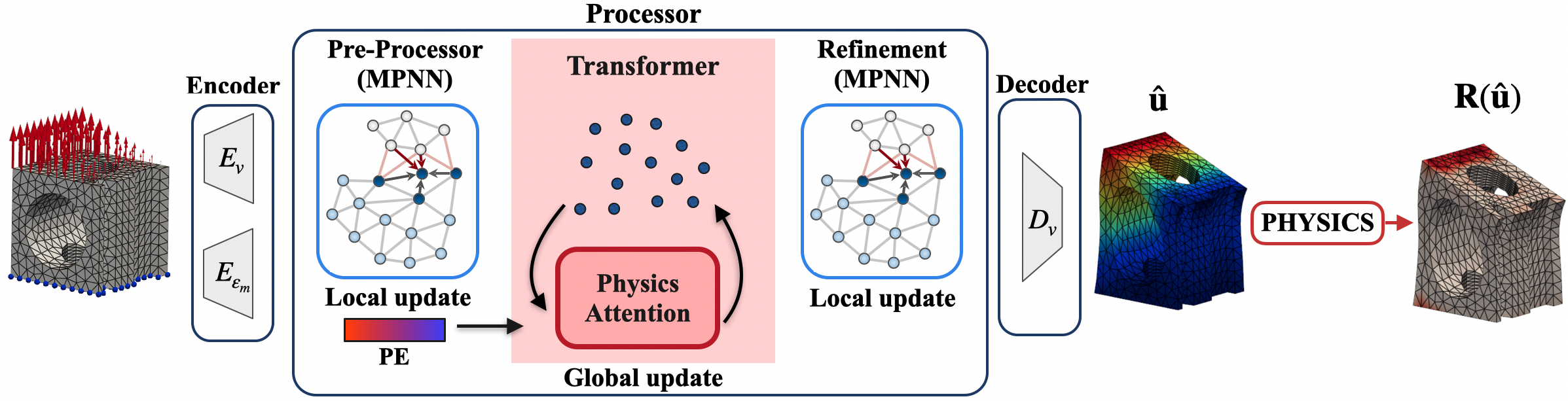}
        \caption{Illustration of MGN-T architecture and physics post-processing.}
        \label{fig:MGNT}
\end{figure}

The MeshGraphnet-Transformer (MGN-T)  architecture \cite{iparraguirre2026MGNT} follows an encoder-processor-decoder framework, combining graph based inductive biases with global attention to efficiently model complex solid mechanics problems (see Figure \ref{fig:MGNT}). The domain is represented by a graph $G(V, E)$ from which node and edge attributes ($\epsilon_V$, $\epsilon_{EM}$) are derived to encode spatial and physical information such as boundary conditions and mesh connectivity. This allows the network to naturally adapt to different geometries and varying loading conditions. Node attributes are defined as $\epsilon_V = (\bs{n}_i, \bar{\bs{u}}_i)$, where $\bs{n}_i$ is the node type and $\bar{\bs{u}}_i$ {the Dirichlet} prescribed displacements, whereas edge attributes, $\epsilon_{EM} = (\bs{X}_i-\bs{X}_j, ||\bs{X}_i-\bs{X}_j||)$, are relative distances between connected nodes in the reference configuration $\bs{X}$.  MGN-T exploits message-passing neural networks (MPNN) blocks to capture local geometric interactions and enforce mesh connectivity, particularly useful for unstructured {adaptive} meshes, updating node and edge attributes in the latent space. The pre-processor absorbs boundary conditions and local interactions before the global update, while the refinement block reinforces geometric consistency. 

In order to overcome under-reaching and successfully model long-range dependencies, the local MPNN are complemented with a Transformer processor based on physics-attention. A slicing mechanism projects the original mesh of {$n_{\mathrm{nod}}$} nodes onto {$P$} physics tokens in the latent space ({$n_{\mathrm{nod}}\gg P$}). To enable sharper projections a learnable temperature $\tau$ dynamically adjusts the weights $\bs{w}$ to project each node to the $P$ tokens. In addition, {the Gumbel-Softmax} reparametrisation trick is used to sample $\bs{w}$. A 4-head multi-head attention {mechanism} over the physical tokens $P$ is used as physics-attention to update the tokens in the latent space. Once updated, the physical tokens are projected back to the mesh nodes. By combining local updates with global updates, MGN-T provides an efficient and robust scalable mesh-based architecture for solid mechanics simulation in many query problems with varying geometry and boundary conditions. In this work, we employ this architecture to predict the displacement field solution to the PDE-constraint problem for 3D complex geometries.

\subsection{Supervision strategy}

To clarify the performance of our data-free approach, we contrast the proposed physics-informed loss
function (Section \ref{sec:physics_loss}) against a standard data-driven supervised loss (Section \ref{sec:data-driven}).

\subsubsection{Physics-informed loss}
\label{sec:physics_loss}

Based on the weak formulation of solid mechanics, the proposed physics
loss $\mathcal{L}_{\text{phys}}$ couples two nodal-level metrics—the equilibrium force
residual $\mathcal{L}_R$ and the boundary condition penalty $\mathcal{L}_{\text{bc}}$---with
an element-level inverted penalty $\mathcal{L}_{\mathrm{neg}J}$:
\begin{equation}
\label{eq:physics_loss}
\mathcal{L}_{\text{phys}} = \log\left(1 + \mathcal{L}_{R}\right) + \lambda_{1}\mathcal{L}_{\mathrm{neg}J} + \lambda_{2}\mathcal{L}_{\text{bc}},
\end{equation}
where $\lambda_1=\lambda_2=0.1$ are scalar weighting hyperparameters.
{Let $Q$ denote the family of queries of a given example and $Q_{\mathrm{train}}\subset Q$ the subset used for training; how $Q$ is sampled is described for each example in Section~\ref{sec:examples}, and its random partition into training, validation and test queries in Section~\ref{sec:experiments}. The optimisation problem for training the data-free neural operator is
\begin{equation*}
\label{eq:training}
\bs{\theta}^* = \mathop{\mathrm{arg\,min}}_{\bs{\theta}}\ \frac{1}{|Q_{\mathrm{train}}|}\sum_{q\in Q_{\mathrm{train}}} \mathcal{L}_{\text{phys}}\big(\hat{\bs{u}}_{\bs{\theta}}(q);q\big),
\end{equation*}
where $|Q_{\mathrm{train}}|$ is the number of training queries and $\hat{\bs{u}}_{\bs{\theta}}(q)$ is given by (\ref{eq:prediction}).\par}

All sub-terms in the loss are directly derived from the {predicted nodal displacements $\hat{\bs{u}}_{\bs{\theta}}$ and the finite element shape functions $N^e_a$} (see Figure \ref{fig:pipeline_physics_calculation}). The sub-terms are described below, with detailed pseudo-code in Appendix \ref{ap:pseudo-code}:
\begin{itemize}
    \item \textbf{Residual forces ($\mathcal{L}_R$):} {Mean of the squared out-of-equilibrium force components over the free degrees of freedom $\mathcal{D}_{\mathrm{free}}$}
    \begin{equation*}
        {\mathcal{L}_{R} = \frac{1}{{|\mathcal{D}_{\mathrm{free}}|}} \sum_{(i,k)\in\mathcal{D}_{\mathrm{free}}} R_{ik}(\hat{\bs{u}}_{\bs{\theta}})^2},
    \end{equation*}
    {where $R_{ik}$ are the components of the nodal residual defined in Eq. (\ref{eq:residual}). 
    In early training $\mathcal{L}_R$ exceeds its converged value by many orders of magnitude. We therefore minimise $\log(1+\mathcal{L}_R)$, a strictly increasing 
    transformation that vanishes only at $\mathcal{L}_R=0$; consequently, it has exactly the same minimisers as $\mathcal{L}_R$. 
    Its gradient, $\nabla_{\bs{\theta}}\mathcal{L}_R/(1+\mathcal{L}_R)$, damps the very large gradients of early training, 
    while for $\mathcal{L}_R\ll 1$ we have $\log(1+\mathcal{L}_R)\approx\mathcal{L}_R$ and the original loss is recovered. 
    The offset of one is what ensures both properties: without it, $\log\mathcal{L}_R$ would be unbounded below as $\mathcal{L}_R\to 0$, 
    with gradient $\nabla_{\bs{\theta}}\mathcal{L}_R/\mathcal{L}_R$. This choice is empirical and aims at balancing the residual term against 
    the penalty terms; it does not alter the minimisers, nor does it improve the conditioning of the underlying problem or the relation between residual 
    and error discussed in Sections~\ref{sec:future}.}

    \item \textbf{Boundary conditions ($\mathcal{L}_{\text{bc}}$):} Enforces kinematic constraints
    by minimising the mean squared error across all prescribed displacement degrees of
    freedom, {$n_{\mathrm{bc}}$}:
    \begin{equation*}
        \mathcal{L}_{\text{bc}} = \frac{1}{{n_{\mathrm{bc}}}}\sum_{i=1}^{{n_{\mathrm{bc}}}} \|\bar{\bs{u}}_i - \hat{\bs{u}}_i\|_2^2.
    \end{equation*}

    \item \textbf{Negative Jacobians ($\mathcal{L}_{\mathrm{neg}J}$):}
    Penalises non-physical element inversion during large deformations.
    It explicitly filters out valid, non-negative element volumes:
    \begin{equation*}
        \mathcal{L}_{\mathrm{neg}J} = \sum_{e=1}^{{n_{\mathrm{el}}}} \|\max(0, -J^{e})\|_2^2,
    \end{equation*}
    where {$J^{e}=\det\bs{F}^h|_{\Omega_e}$} is the determinant of the deformation gradient in element $e$, and {$n_{\mathrm{el}}$} is the total number of elements. {For numerical stability, $J^e$ is bounded below by $0.1$ in the constitutive evaluation. In strongly compressed or inverted elements the computed stress, and hence the residual, is therefore not the exact finite element one; $\mathcal{L}_{\mathrm{neg}J}$ drives the prediction away from this regime.}
\end{itemize}

\subsubsection{Data-Driven Loss}
\label{sec:data-driven}

For comparison, a standard data-driven framework is also considered, that minimises the mean-squared discrepancy between
 the predicted {$\hat{\bs{u}}_i$ and reference $\bs{u}^\star_i$} displacements across all {$n_{\mathrm{nod}}$} nodes:
\begin{equation}
\label{eq:data_loss}
\mathcal{L}_{\text{data}} = \frac{1}{{n_{\mathrm{nod}}}} \sum_{i=1}^{{n_{\mathrm{nod}}}}\|{\bs{u}^\star_i} - \hat{\bs{u}}_i\|^2.
\end{equation}
{Being a Euclidean norm of nodal values, (\ref{eq:data_loss}) is a discrete $L^2$ norm of the displacement error. Since the reference solutions satisfy equilibrium, (\ref{eq:data_loss}) measures the distance to the equilibrium solution directly, but it requires a finite element solution for every training query.\par}

\section{Numerical Examples}\label{sec:examples}

Three numerical examples are presented: an homogeneous stress problem on a cube, and two problems on intricate geometries: a perforated cube and a plate. For all examples, a set of queries $Q$ defines the regression range in terms of loading scenarios and geometries.

To assess the performance of the proposed approach, we first validate the model against the patch test on a cube. As comparisons with commercial solvers are often difficult to make due to their proprietary and closed nature, the patch test allows us to compare between data-driven and physics-informed approaches. The remaining examples are solved and directly evaluated through the {equilibrium check}, see Eq.  (\ref{sec:appendix_criteria}).

\subsection{Homogeneous traction on a cube}

A 3D neo-Hookean cube under pure traction-compression states is considered. A total of 100 different loading cases comprise the queries $Q$, all sharing the same geometry and Dirichlet boundary conditions. The bottom side is constrained along $z$-axis, while the left and right sides along $x$-axis and $y$-axis, respectively. The top surface has a prescribed displacement sampled from $u_z \in [-10, 10]$. The remaining sides are free.

\begin{figure}[h!]
    \centering
    \begin{subfigure}{0.24\textwidth}
        \centering
        \includegraphics[width=\linewidth]{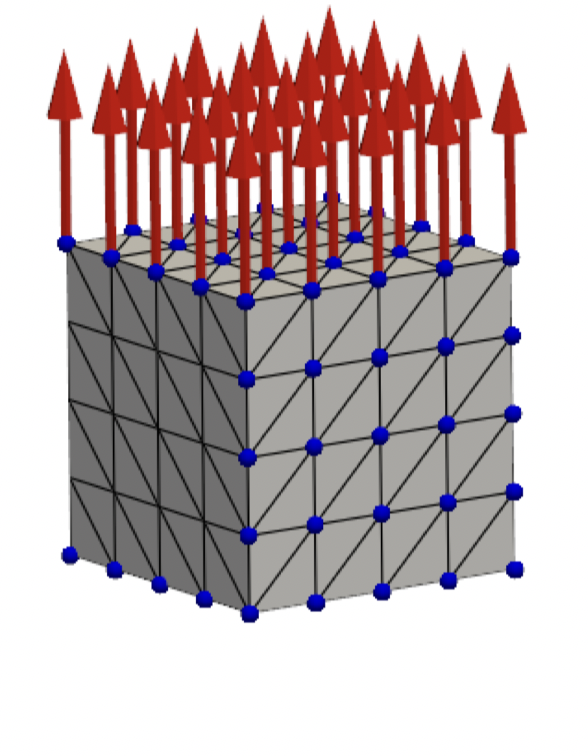}
        \caption{Test query}
        \label{fig:sub1}
    \end{subfigure}
    \hspace{0.04\textwidth}
    \begin{subfigure}{0.24\textwidth}
        \centering
        \includegraphics[width=\linewidth]{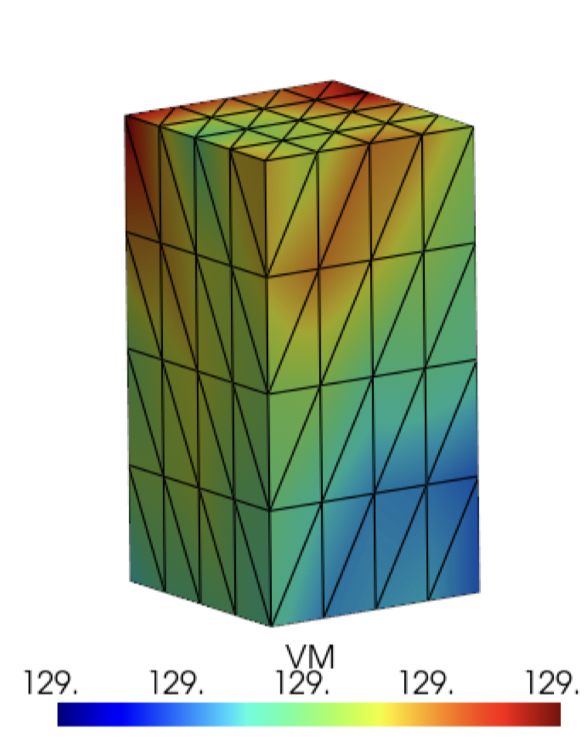}
        \caption{$\sigma_{\text{VM}}(\hat{\bs{u}})$}
        \label{fig:sub2}
    \end{subfigure}
    \hspace{0.04\textwidth}
    \begin{subfigure}{0.24\textwidth}
        \centering
        \includegraphics[width=\linewidth]{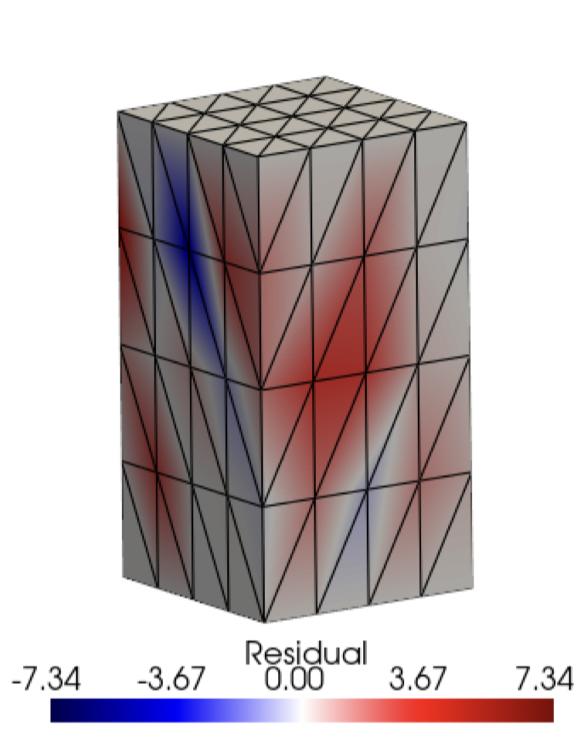}
        \caption{$\bs{R}(\hat{\bs{u}})$}
        \label{fig:sub3}
    \end{subfigure}
    \caption{Test prediction for unseen loading case for the homogeneous traction on a cube. a) problem statement, where blue nodes represent constrained sides and red arrows prescribed {displacements}. b) $\sigma_{\text{vM}}$ derived from predicted $\hat{\bs{u}}$.  c) Corresponding nodal residuals.}
    \label{fig:patch_test}
\end{figure}

The cube has dimensions $20\times 20\times 20$ units and is discretised using 156 nodes and tetrahedral elements (see Figure \ref{fig:patch_test}).

\subsection{Intricate geometries}

Two geometries are described: a perforated cube and a plate. The perforated cube consists of a 3D cube with several extrusions and perforations, whereas the plate represents an industrial component geometry.

\begin{figure}[htbp]
    \includegraphics[width=\linewidth]{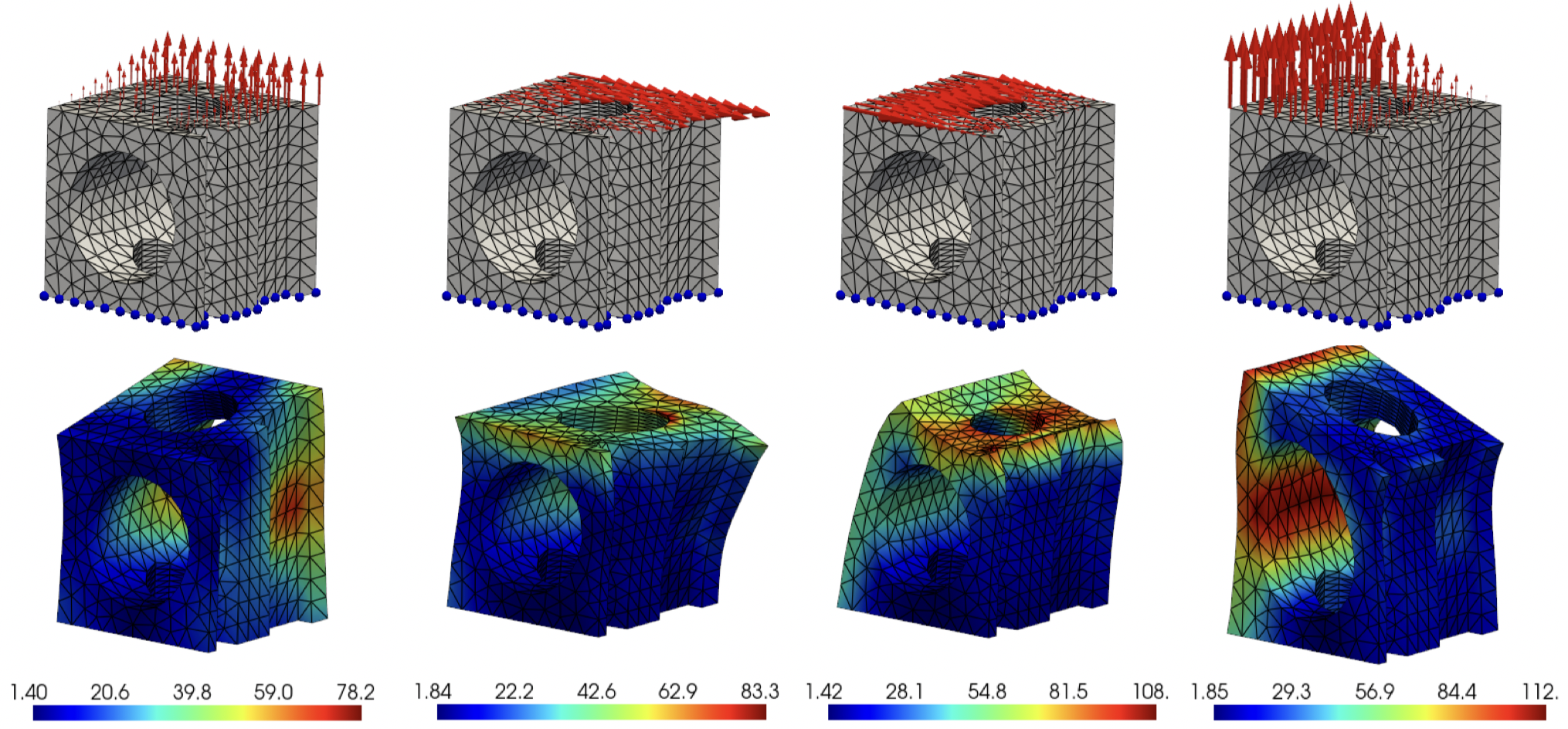}
    \caption{Test predictions for unseen loading cases for perforated cube geometry. Top: Prescribed displacements. Bottom: colour map of the predicted von Mises stress $\sigma_{\text{vM}}$. Blue dots indicate prescribed displacements.}
    \label{fig:test_cube_perfored}
\end{figure}

These are also discretised using tetrahedral elements with an adaptive mesh, consisting of 1K and 2.5K nodes, respectively.

Each query consists of a different loading curve comprising uniform imposed displacements and a ramp along different axes, with amplitude in $ [-10,10]$. A total of 200 queries are defined; see Figure \ref{fig:test_cube_perfored}.

\begin{figure}[h!]
\centering
   \begin{subfigure}[b]{0.65\textwidth}
       \centering
        \includegraphics[width=\linewidth]{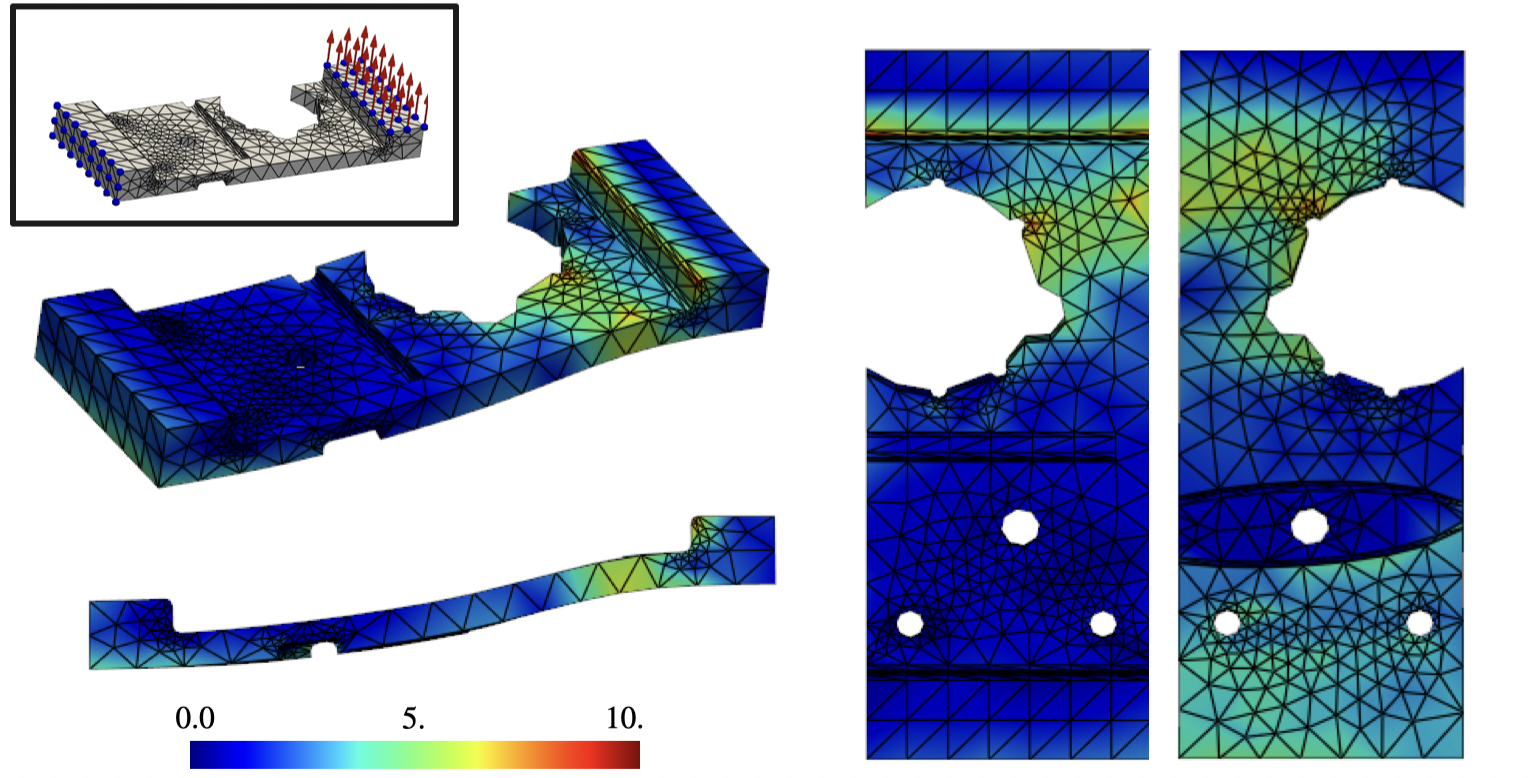}
        \caption{}
        \label{fig:plate_vmises}
    \end{subfigure}
    \vfill
    \begin{subfigure}[b]{0.65\textwidth}
       \centering
        \includegraphics[width=\linewidth]{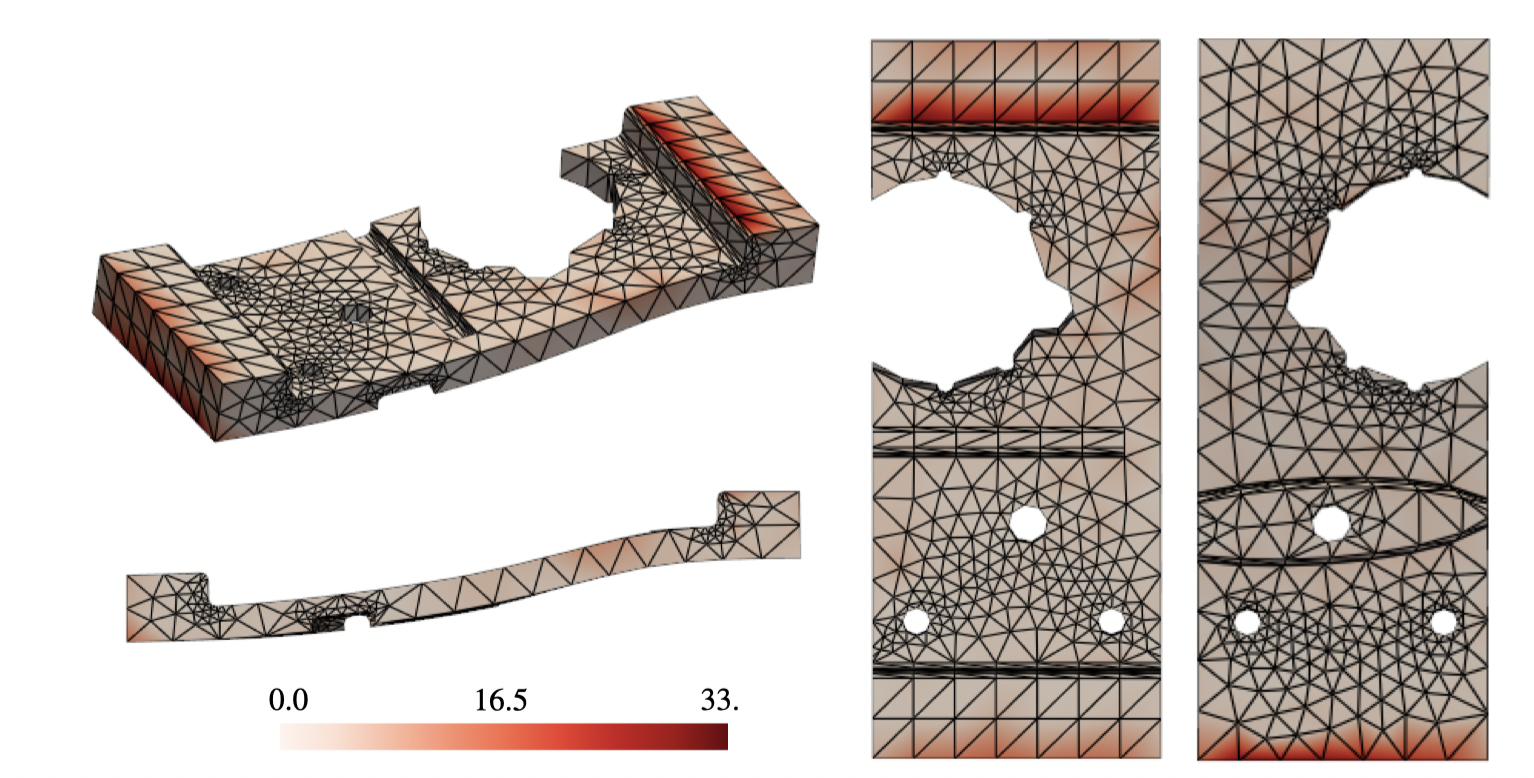}
        \caption{}
        \label{fig:plate_residual}
    \end{subfigure}
    \caption{Test predictions for unseen loading cases for the plate geometry. (a) Colour maps represent von Mises stress $\sigma_{\text{vM}}$. (b)  Norm of the nodal residuals $\|\bs{R}\|$.}
    \label{fig:NLGEOM}
\end{figure}

\section{Experiments}\label{sec:experiments}

The neo-Hookean material is described by Lam\'e's constants $\lambda = 40$ and $\mu = 0.01$. Models are trained with the Adam optimiser and a cosine-annealed learning scheduler from $lr=10^{-4}$ to $lr=10^{-6}$ on a single RTX 4090 GPU \cite{kingma2017adam, loshchilov2017}. {The same MGN-T configuration is used in all examples: two MPNN blocks, one in the pre-processor and one in the refinement stage (Section~\ref{sec:meth}), and a latent dimension of 64 for all node, edge and token features. Only the number of physics tokens of the global processor changes with the size of the problem: $P=8$ for the homogeneous cube, $P=16$ for the perforated cube and $P=32$ for the plate.}
{To assess the performance of the method, we use the equilibrium check $r_{\max}$ of Eq.~(\ref{sec:appendix_criteria}) for physical consistency, and the relative root-mean-square error of the nodal displacements, $\mathrm{RRMSE}(\hat{\bs{u}})=\|\hat{\bs{u}}-\bs{u}^\star\|_2/\|\bs{u}^\star\|_2$, for the similarity between prediction and reference solution. The reference is the analytical solution for the homogeneous cube and a finite element solution on the same mesh otherwise.}

To prevent over-fitting, the family of query problems $Q$ associated with each numerical example is randomly partitioned into training, validation, and test sets containing 70\%, 15\%, and 15\% of the queries, respectively. {Learning curves are monitored on the validation queries, and all other reported metrics are computed on the test queries unless stated otherwise.}

\subsection{Physics-informed vs. data-driven learning}

This section compares the results obtained when training our architecture using a loss function informed by physics or one based solely on data. Both strategies are compared on the homogeneous traction cube problem.

\begin{figure}[h!]
\centering
\begin{tikzpicture}
\begin{axis}[
    ymode=log,
    ylabel={Error},
    xtick={1.5, 3.5},
    xticklabels={$r_{\text{max}}$, RRMSE($ \hat{\bs{u}}$)},
    width=7cm,
    height=5cm,
    grid=both,
    boxplot/draw direction=y,
    legend style={
        at={(0.8,0.7)},
        anchor=south,
        legend columns=1,
        draw=none
    },
]
\addlegendimage{area legend, fill=blue!50, draw=blue}
\addlegendentry{$\mathcal{L}_{\text{data}}$}

\addlegendimage{area legend, fill=red!50, draw=red}
\addlegendentry{$\mathcal{L}_{\text{phys}}$}
\addplot+[
    boxplot prepared={
        lower whisker=0.0009712986602160265,
        lower quartile= 0.001444362516861291,
        median=0.003028611867194647,
        upper quartile=0.00442139104344465,
        upper whisker=0.00489921159227904
    },
    fill=blue!30,
    draw=black
] coordinates {(1,0)};

\addplot+[
    boxplot prepared={
        lower whisker=0.00026742923225003454,
        lower quartile=0.00035811797503361133,
        median=0.000625159720627817,
        upper quartile=0.001245270423559388,
        upper whisker=0.0017670321043816977
    },
    fill=red!30,
    draw=black
] coordinates {(1,0)};


\addplot+[
    boxplot prepared={
        lower whisker=3.6685373243300464e-07,
        lower quartile=7.921795135310933e-07,
        median=2.5205372428820633e-06,
        upper quartile=4.114119336111501e-06,
        upper whisker=7.808253332763882e-06
    },
    fill=blue!30,
    draw=black
] coordinates {(2,0)};

\addplot+[
    boxplot prepared={
        lower whisker=3.756125031023144e-06,
        lower quartile=6.663022425518882e-06,
        median=1.2560365616600734e-05,
        upper quartile=1.8522488690545753e-05,
        upper whisker=2.3763250471600897e-05
    },
    fill=red!30,
    draw=black
] coordinates {(2,0)};

\end{axis}
\end{tikzpicture}

\caption{Comparison of the {equilibrium check $r_{\max}$} and RRMSE displacement errors for unseen test loading cases.
{Blue boxes correspond to the model trained on $\mathcal{L}_{\text{data}}$ and red ones to the model trained on $\mathcal{L}_{\text{phys}}$.}}
\label{fig:boxplot}
\end{figure}

Two training strategies are conducted. First, the proposed physics-informed learning strategy, where the model is trained on the physics loss $\mathcal{L}_{\text{phys}}$, requires no data. Second, a supervised learning
approach, in which the model is trained on the $\mathcal{L}_{\text{data}}$, data is required. As shown in Figure \ref{fig:boxplot}, the strategies perform better on their own metric of training (physics-informed learning has better force residual errors than supervised learning,
vice versa for the data-driven errors), as expected. Nevertheless, achieving low similarity errors does not guarantee physical consistency, showing that data loss can rapidly yield a close-to-solution answer but might fail to fulfil the physical principles. Notice that this scenario is ideal for data-driven training, as the data has no computational noise since it comes from the analytical solution.

\begin{figure}[h!]
    \centering
    \includegraphics[width=0.9\linewidth]{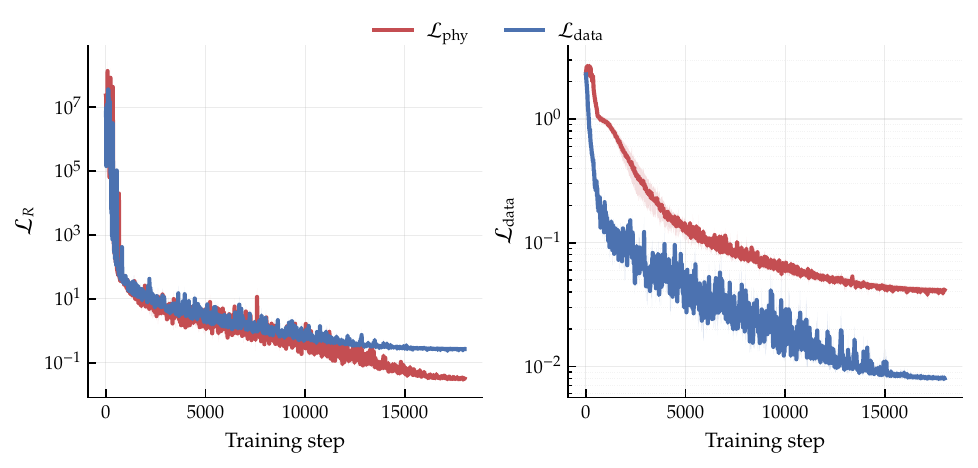}
    \caption{Comparison of the physics-informed strategy (red) vs the data-driven (blue) strategy, monitoring {errors on the validation queries}. Evolution during training for three different seed initialisations; the shaded area represents the standard error.  Left: force balance error $\mathcal{L}_R$. Right: data similarity error $\mathcal{L}_{\text{data}}$.}
    \label{fig:learning_curves}
\end{figure}

Unlike the data-driven setting, where the equilibrium solution $\bs{u}^\star$ is known, 
the physics-informed approach jointly optimises the predicted solution $\hat{\bs{u}}_{\bs{\theta}}$, 
constrained to satisfy equilibrium for every training query, and the network parameters $\bs{\theta}$. 
{Figure~\ref{fig:learning_curves} shows that, for the same number of iterations, physics-informed training reaches lower 
force residuals, whereas data-driven training reaches lower displacement errors. \par}

\begin{figure}[htbp]
    \centering
    \includegraphics[width=\linewidth]{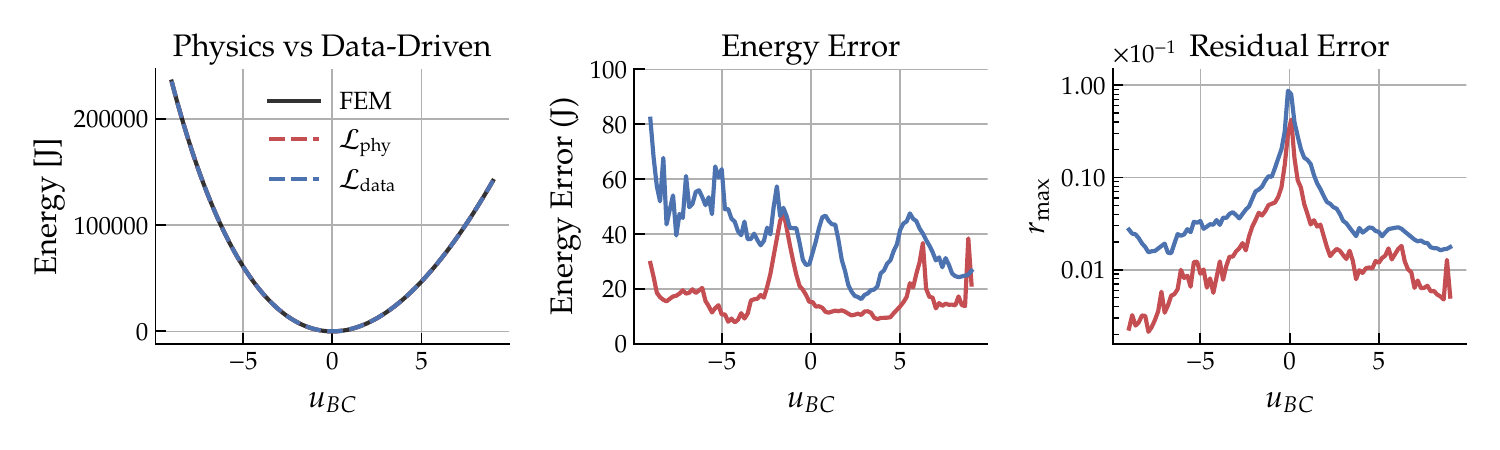}
        \caption{Comparison at inference for continuous input $\bar{u}\in [-10, 10]$.
        From left to right: Total potential energy for given displacement, energy error and {equilibrium check $r_{\max}$}.}
        \label{fig:energy}
\end{figure}

Since the proposed model is trained as a neural operator, it can be evaluated for arbitrary prescribed displacements
within the training domain, enabling continuous dense loading input $\bar{u}\in [-10, 10]$. Figure \ref{fig:energy} 
compares the performance of both strategies in terms of energy and errors. The non-linearity and non-symmetry 
in traction-compression of the neo-Hookean material are captured by both models; however, physics-informed 
strategy achieves better convergence and energy errors. {This dense sweep contains both training and unseen 
boundary values, and the reference is the analytical solution.} {The total potential energy (\ref{eq:energy}) is computed at element level: since $\bs{F}^h$ is constant in each linear tetrahedron, the strain energy density $\Psi$ of Section~\ref{sec:neo_hook} is evaluated once per element and integrated exactly over its reference volume, $\Pi(\hat{\bs{u}})=\sum_{e=1}^{n_{\mathrm{el}}}\Psi(\bs{F}^e)\,V^e$. No external work term appears, since the loading is applied only through prescribed displacements.}

{In summary, the physics-informed loss requires no finite element data, its value can be interpreted directly as an equilibrium check, and it can be evaluated and further minimised at inference for queries never seen in training. The data loss measures the error directly, but requires a finite element solution for every training query and provides no indicator of physical consistency at inference. Although residual losses are known to be worse conditioned than data losses (Section~\ref{sec:data-driven}), both losses converge at a similar rate on the validation queries of this example (Figure~\ref{fig:learning_curves}). We attribute this to two factors. First, the displacement field is not optimised directly but through the parameters of a network, $\hat{\bs{u}}_{\bs{\theta}}$, whose non-convex parametrisation governs the optimisation landscape for both losses. Second, the mesh of this example is small, which limits the condition number of the stiffness matrix. Whether this behaviour persists for larger and finer meshes remains to be assessed.\par}

\subsection{Intricate geometries}

In this section, we assess the generalisation capabilities to unseen loading cases and different, complex geometries. As no data is required, 
the goal is not just to solve all the training cases, but also to provide physically consistent solutions at inference. 
For instance, Figure \ref{fig:test_cube_perfored}  shows predictions for unseen loading test cases for the perforated cube example, and 
Figure~\ref{fig:NLGEOM} presents the prediction for an unseen loading case on the plate geometry.
{Once trained, a single forward pass of the network takes $2.2\times10^{-3}$\,s for the perforated cube and $2\times10^{-2}$\,s for the plate on a single RTX 4090 GPU, and directly returns the displacement field of the query. By contrast, a finite element solver must apply the prescribed displacements incrementally, since large deformations are involved, and solve each load increment with several Newton--Raphson iterations, the number of increments depending on the loading conditions.}

\begin{figure}[h!]
    \includegraphics[width=\linewidth]{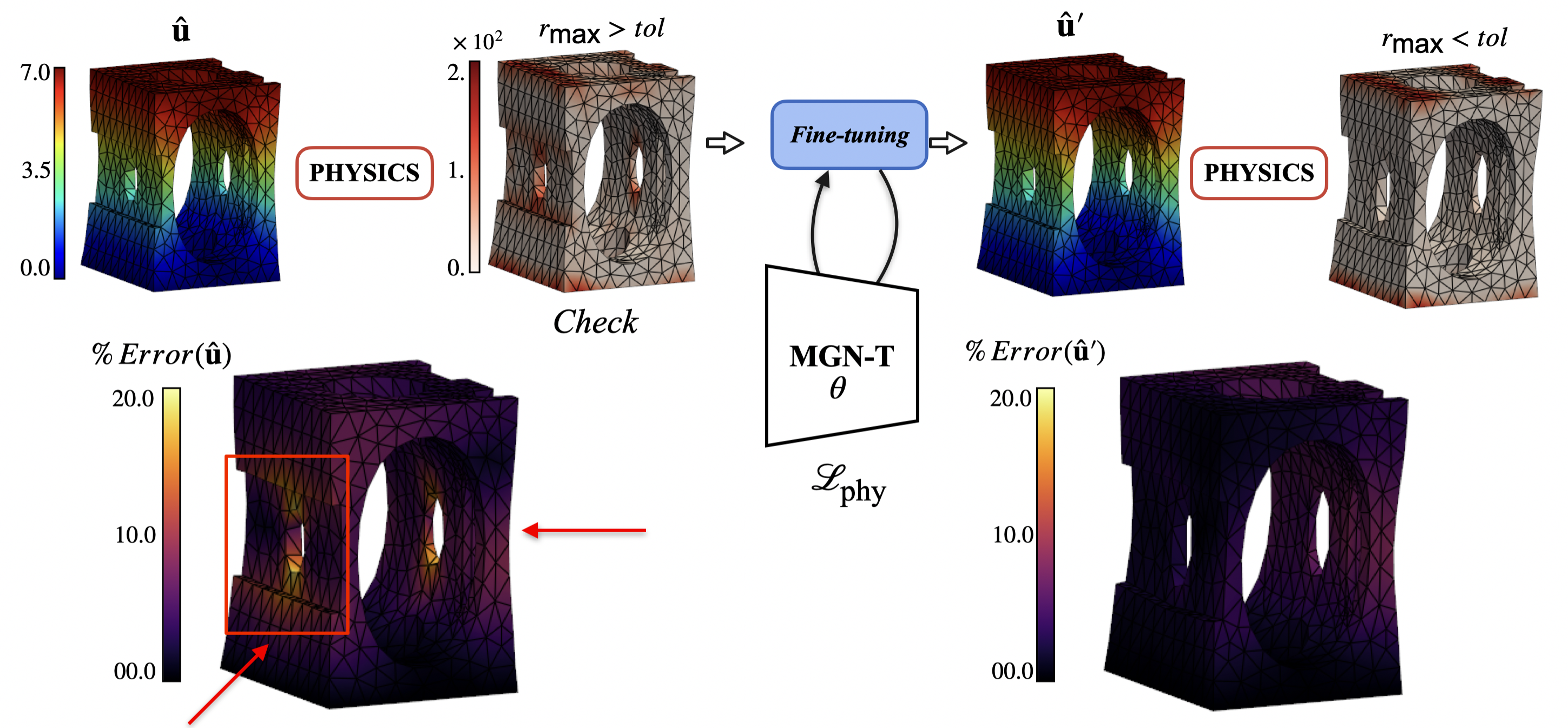}
    \caption{{Test-time fine-tuning} example for an unseen geometry. The first two figures, top left, show the displacement field $\bs{\hat{u}}$ and the residuals $r_{\text{max}}$ obtained by direct inference from the new geometry, which results from a modification of the training geometry. 
    {If the equilibrium check fails, the network is fine-tuned on this query with the physics loss (top right: fine-tuned displacement field $\bs{\hat{u}'}$, whose residuals are below the tolerance).} The bottom row shows the error fields in displacement relative to the reference finite element solution without {fine-tuning} (left) and with {fine-tuning} (right).
    Highlighted with red arrows are the major geometry changes with respect to train geometry.}
    \label{fig:test_refinement}
\end{figure}

To assess the capability to generalise to unseen geometries, a new geometry is considered, consisting of an additional through-hole and a rectangular 
extrusion on the left side for the perforated Cube. For comparison, the displacement field is obtained through finite element simulation. 
Figure \ref{fig:test_refinement} top, left, shows the displacement field together with the corresponding force residual. Despite the geometric modification, 
the model achieves an overall relative RMSE below 5\% in displacements, with the largest errors and force residuals concentrated around the regions most 
affected by the geometric changes.

{As shown, the architecture handles varying mesh resolutions and complex topologies. These results show that the proposed approach provides predictions 
for unseen loading cases and geometries that satisfy the equilibrium check, either directly or after test-time fine-tuning.}

\subsection{{Test-time fine-tuning}}

{During inference, the equilibrium check (\ref{sec:appendix_criteria}) is evaluated for each prediction at negligible cost. 
Whenever it fails, the network is fine-tuned on that query, or on a batch of queries, by minimising the same physics loss until the check is 
met (see Figure \ref{fig:test_refinement}). We use the term \emph{fine-tuning} rather than \emph{refinement} to avoid confusion with mesh refinement: 
the mesh is unchanged and only the network parameters are updated. In the example of Figure~\ref{fig:test_refinement}, the zero-shot prediction exhibits 
a force residual above the tolerance. Fine-tuning reduces the force residuals below the tolerance and, in this example, 
also reduces the displacement error with respect to the finite element reference from about 5\% to below 3\% relative RMSE. 
For a batch of new queries consisting of unseen geometries and loading cases, Figure \ref{fig:bolxplot_nlgeom} shows that the model meets 
the equilibrium check after fine-tuning in approximately 10\% of the original training iterations. This procedure can be fully automated once 
the user has selected the residual tolerance, or it can be left to the user's discretion.\par}

\begin{figure}[h!]
\centering
\begin{subfigure}[t]{0.48\textwidth}
\centering
\begin{tikzpicture}
\begin{axis}[
    ymode=log,
    ymin=2e-4,
    ymax=5e-2,
    ylabel={$r_{\max}$},
    xtick={1,2},
    xticklabels={train, test},
    width=\textwidth,
    height=4cm,
    grid=both,
    boxplot/draw direction=y,
]
\addplot+[
    boxplot prepared={
        lower whisker=0.0006194284,
        lower quartile=0.0024242516,
        median=0.0041468799,
        upper quartile=0.0060369469,
        upper whisker=0.0113450133
    },
    fill=purple!60,
    draw=black
] coordinates {};

\addplot+[
    boxplot prepared={
        lower whisker=0.0006267531,
        lower quartile=0.0031879235,
        median=0.0056335257,
        upper quartile=0.0119014161,
        upper whisker=0.0233690140
    },
    fill=purple!30,
    draw=black
] coordinates {};

\end{axis}
\end{tikzpicture}
\caption{Seen geometry}
\end{subfigure}
\hfill
\begin{subfigure}[t]{0.48\textwidth}
\centering
\begin{tikzpicture}
\begin{axis}[
    ymode=log,
    ymin=2e-4,
    ymax=5e-2,
    ylabel={$r_{\max}$},
    xtick={1,2},
    xticklabels={zero-shot, {fine-tuned}},
    width=\textwidth,
    height=4cm,
    grid=both,
    boxplot/draw direction=y,
]
\addplot+[
    boxplot prepared={
        lower whisker=0.006401997789537443,
        lower quartile= 0.0082804,
        median=0.010804,
        upper quartile=0.0126998656,
        upper whisker=0.0175630790
    },
    fill=orange!60,
    draw=black
] coordinates {(1,0)};
\draw[dashed, thick]
(axis cs:0,1e-2) -- (axis cs:3,1e-2);
\addplot+[
    boxplot prepared={
        lower whisker=0.00289548,
        lower quartile= 0.0047731104,
        median=0.005601796,
        upper quartile=0.0063224,
        upper whisker=0.00839411638124257
    },
    fill=orange!30,
    draw=black
] coordinates {(2,0)};
\end{axis}
\end{tikzpicture}
\caption{Unseen geometry}
\end{subfigure}
\caption{{Equilibrium check} $r_{\max}$ for the perforated cube example. The dashed line represents {the tolerance}.
 a) Seen geometry: train and test queries b) Unseen geometry: {zero-shot prediction and after test-time fine-tuning.}}
\label{fig:bolxplot_nlgeom}
\end{figure}

Furthermore, the force residual may serve as a heuristic indicator of whether a query lies within the regime covered by training, as observed for the unseen geometry in Figure~\ref{fig:bolxplot_nlgeom}, whose zero-shot predictions fail the check. 
{Using residuals to assess accuracy across the parameter space is, however, delicate in general \cite{bachmayr2024variationally}, and we do not claim this as a certified indicator.\par}
Our approach also enables continuous learning, as new queries and geometries can be progressively incorporated to extend the regression domain.

\subsection{{Ablation: negative-Jacobian penalty}}
\label{sec:ablation_negj}

\begin{figure}[htbp]
    \centering
    \includegraphics[width=\linewidth]{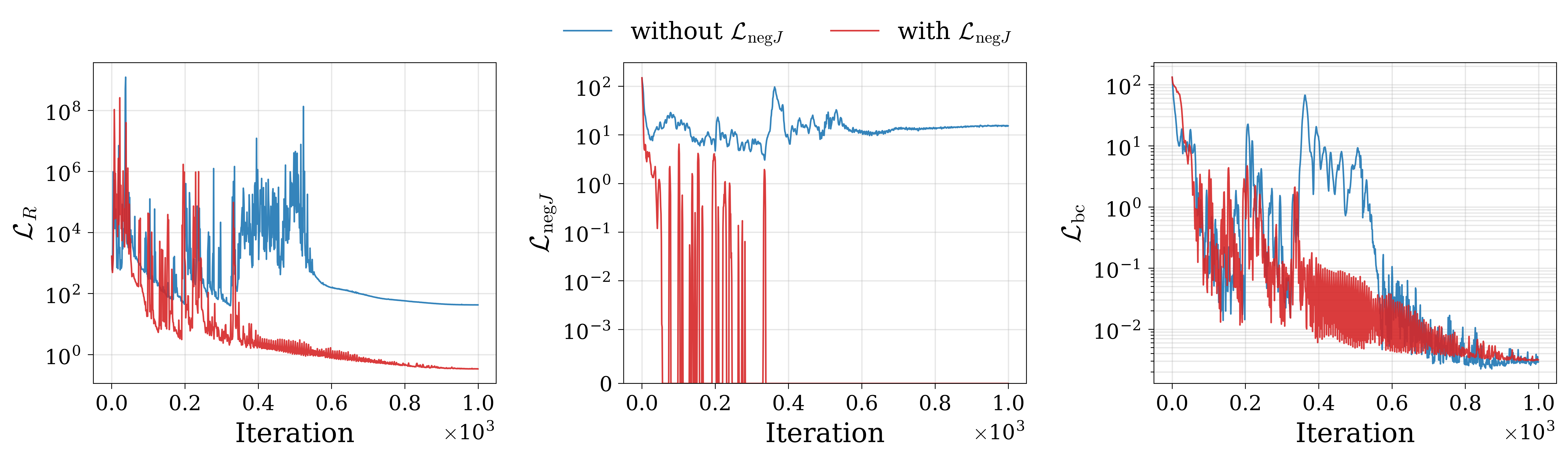}
    \caption{{Ablation of the negative-Jacobian penalty on the plate example: evolution of $\mathcal{L}_R$, $\mathcal{L}_{\mathrm{neg}J}$ and $\mathcal{L}_{\mathrm{bc}}$ during training with (red) and without (blue) $\mathcal{L}_{\mathrm{neg}J}$ in the loss. Solid lines: training queries; dashed lines with markers: validation queries. $\mathcal{L}_{\mathrm{neg}J}$ is shown on a symmetric logarithmic scale, linear below $10^{-3}$, since it reaches exactly zero.}}
    \label{fig:ablation_negj}
\end{figure}

\begin{figure}[htbp]
    \centering
    \includegraphics[width=\linewidth]{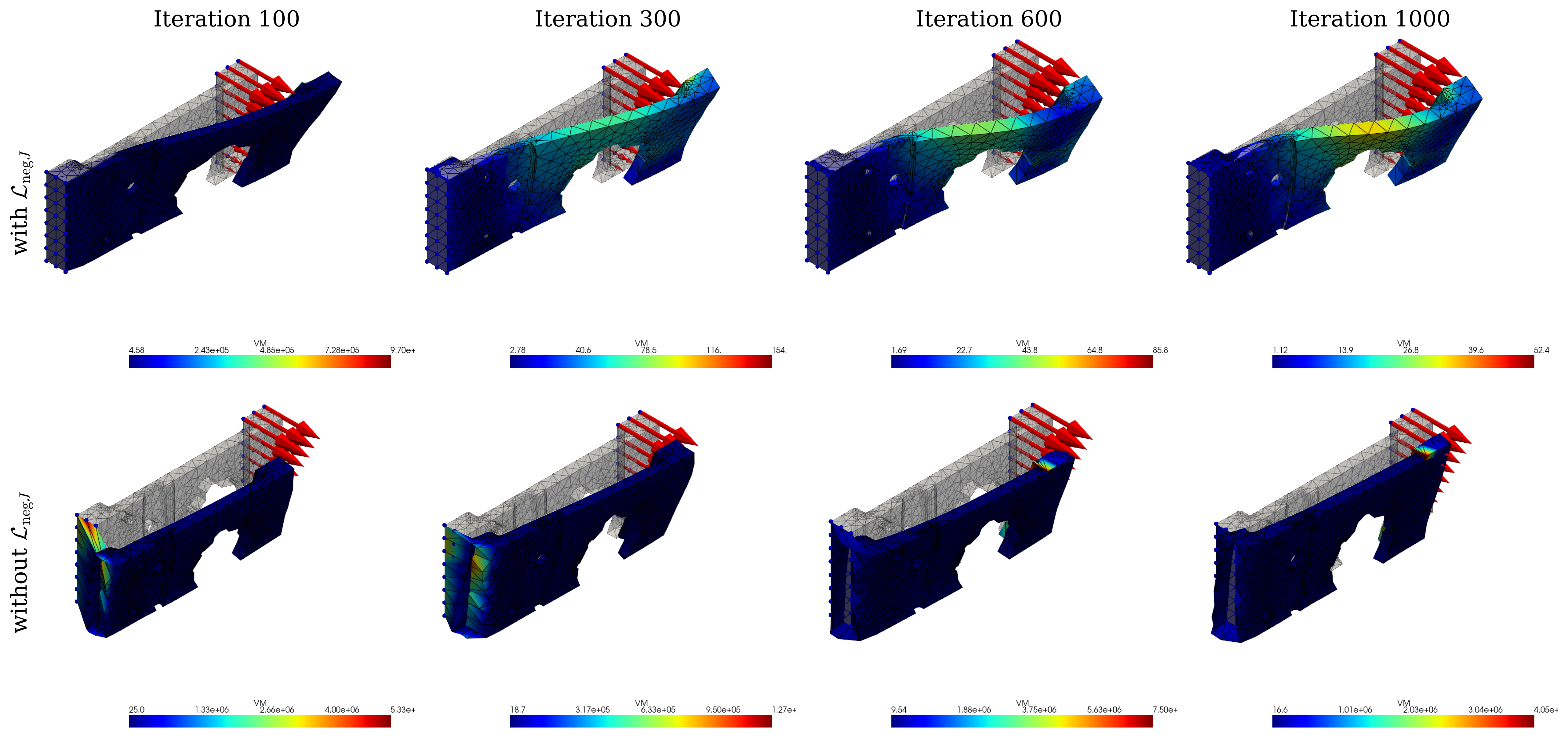}
    \caption{{Predicted deformed configuration and von Mises stress for one query of the plate example at iterations 100, 300, 600 and 1000, trained with (top) and without (bottom) $\mathcal{L}_{\mathrm{neg}J}$. Note that each panel has its own colour scale.}}
    \label{fig:ablation_negj_snapshots}
\end{figure}

{To assess the role of the negative-Jacobian penalty, the model is trained on the plate example with and without 
$\mathcal{L}_{\mathrm{neg}J}$ ($\lambda_1=0$), keeping all other settings unchanged (Figures~\ref{fig:ablation_negj} and~\ref{fig:ablation_negj_snapshots}). 
With the penalty, inverted elements disappear early in training and $\mathcal{L}_R$ decreases steadily. Without it, the prediction never leaves the 
configurations with inverted elements: $\mathcal{L}_{\mathrm{neg}J}$ never vanishes, $\mathcal{L}_R$ stagnates two orders of magnitude higher, and 
the collapsed elements concentrate spurious stresses, so that the problem is not solved. \par}

\section{{Future work}}
\label{sec:future}

Several extensions are left for future work. Although the formulation is general, the current implementation employs linear tetrahedra with a single integration point, and higher-order elements will be incorporated. Regarding the material, only homogeneous hyperelastic neo-Hookean solids have been studied; the framework will be extended to other constitutive laws, including path-dependent behaviour such as plasticity or damage, and to heterogeneous materials.

{Regarding scalability, the examples presented here involve meshes of up to a few thousand nodes. The MGN-T architecture has been shown to scale to meshes with more than $10^5$ nodes \cite{iparraguirre2026MGNT}, and we expect scalability to be governed mainly by this architectural choice. The proposed physics-informed loss is not expected to be a major limitation: it is computed element-wise, with a cost linear in the number of elements, and requires no linear solves. Its application to industrial-scale meshes will be addressed in future work.\par}

{Regarding the loss, the Euclidean residual used here is neither equivalent to the error nor well conditioned as shown in Robust VPINNs \cite{rojas2024robust}, their proposed dual norm would address both issues and provide an error-equivalent criterion for test-time fine-tuning. Its extension to the parametric, many-query setting \cite{bachmayr2024variationally} will also be considered.\par}

\section{Conclusion}
\label{sec:conclusion}
In this work, we introduced an end-to-end trainable, physics-informed learning framework for 3D solid mechanics problems undergoing large deformations, without requiring training data. Based on the weak formulation of the equilibrium problem, we solve complex geometries for nonlinear solid mechanics problems in the quasi-static regime. The proposed physical loss function guides neural network training without requiring training data, by coupling equilibrium force residuals and kinematic boundary condition enforcement with a negative Jacobian penalty to prevent element inversion under severe deformations.

Furthermore, the use of an architecture based on graph neural networks and message passing lends the resulting method a geometric bias, the importance of which is amply demonstrated by the existing literature on the subject.

{In the homogeneous cube example, physics-informed training reached lower force residuals than data-driven training at a similar computational cost, whereas data-driven training reached lower displacement errors. The observed behaviour results from the combination of the loss, the architecture and the optimisation.
\par}
At inference time, the trained model acts as a continuous neural operator, providing solutions for unseen loading cases and geometries near-instantaneously 
without requiring traditional time-consuming Newton–Raphson iterations. By embedding the finite element equilibrium equations in the loss function, 
every prediction can be checked for equilibrium at inference and, when the check fails, fine-tuned in a few training steps. By combining the fast regression 
capabilities of the MGN-T architecture with the equilibrium equations of the finite element method, this approach takes a step from purely data-driven learning 
towards physically consistent surrogates.

\section*{{Declaration of generative AI and AI-assisted technologies in the writing process}}

{During the preparation of this work, the authors used Claude (Anthropic) to improve the writing, language and spelling of the manuscript. After using this tool, the authors reviewed and edited the content as needed and take full responsibility for the content of the publication.\par}

\section*{Acknowledgements}

The authors acknowledge the support of the Spanish Ministry of Science and Innovation, AEI/10.13039/501100011033 (grant no. PID2023-147373OB-I00). The authors also acknowledge the support of the Ministry for Digital Transformation and the Civil Service, through the ENIA 2022 Chairs for the creation of the university-industry chairs in Artificial Intelligence, through grant TSI-100930-2023-1.

Funded by the European Union. Views and opinions expressed are however those of the author(s) only and do not necessarily reflect those of the European Union or the European Research Council Executive Agency. Neither the European Union nor the granting authority can be held responsible for them.

This work is supported by ERC grant PHYSIA 101264273.


\vfill\pagebreak

\appendix{}

\section{Pseudo-code algorithm}
\label{ap:pseudo-code}

\begin{algorithm}

\caption{Physics-informed loss and equilibrium check for one query, via tensor contraction and scatter-add}
\label{alg:forward}
\begin{algorithmic}[1]
\Require predicted nodal displacements $\hat{\bs{u}}\in\mathbb{R}^{n_{\mathrm{nod}}\times 3}$, prescribed displacements $\bar{\bs{u}}$, Dirichlet mask $\Gamma_D\in\{0,1\}^{n_{\mathrm{nod}}\times 3}$ (per degree of freedom), connectivity $\mathcal{E}\in\mathbb{N}^{n_{\mathrm{el}}\times 4}$, shape-function gradients $\nabla_0 N\in\mathbb{R}^{n_{\mathrm{el}}\times 4\times 3}$, reference volumes $V\in\mathbb{R}^{n_{\mathrm{el}}}$, Lam\'e constants $\lambda,\mu$, force scale $s$, stored $\tilde{q}_{\mathrm{stored}}$
\Ensure losses $\mathcal{L}_{R}$, $\mathcal{L}_{\mathrm{bc}}$, $\mathcal{L}_{\mathrm{neg}J}$ and equilibrium check $r_{\max}$

\Statex \textbf{Boundary conditions}
\State $\mathcal{L}_{\mathrm{bc}} \leftarrow \frac{1}{n_{\mathrm{bc}}}\sum_{i=1}^{n_{\mathrm{bc}}} \|\bar{\bs{u}}_i - \hat{\bs{u}}_i\|_2^2$
\State $\hat{\bs{u}} \leftarrow \Gamma_D\odot\bar{\bs{u}} + (1-\Gamma_D)\odot\hat{\bs{u}}$ \Comment{enforce prescribed DOFs}

\Statex \textbf{Kinematics}
\State $\hat{\bs{u}}^{e} \leftarrow \hat{\bs{u}}[\mathcal{E}]$ \Comment{$(n_{\mathrm{el}},4,3)$}
\State $\bs{H}^{e} \leftarrow \mathrm{einsum}_{eai,eaj\rightarrow eij}(\hat{\bs{u}}^{e},\nabla_0 N)$ \Comment{$(n_{\mathrm{el}},3,3)$}
\State $\bs{F}^{e} \leftarrow \bs{I} + \bs{H}^{e}$, \quad $J^e \leftarrow \det\bs{F}^e$

\Statex \textbf{Negative Jacobians}
\State $\mathcal{L}_{\mathrm{neg}J} \leftarrow \sum_{e=1}^{n_{\mathrm{el}}} \max(0,-J^e)^2$
\State $J^e \leftarrow \max(J^e,\,0.1)$ \Comment{bound for the constitutive evaluation}

\Statex \textbf{Constitutive model (compressible neo-Hookean)}
\State $\bs{C}^e \leftarrow (\bs{F}^e)^{T}\bs{F}^e$
\State $\bs{S}^e \leftarrow \mu\,(\bs{I}-(\bs{C}^e)^{-1}) + \lambda\,\ln(J^e)\,(\bs{C}^e)^{-1}$
\State $\bs{P}^e \leftarrow \bs{F}^e\bs{S}^e$

\Statex \textbf{Internal forces and residual}
\State $\bs{f}^{e}_{\mathrm{int}} \leftarrow V^e\,\mathrm{einsum}_{eai,eji\rightarrow eaj}(\nabla_0 N,\bs{P}^e)$ \Comment{$(n_{\mathrm{el}},4,3)$}
\State $\bs{f}_{\mathrm{int}} \leftarrow \mathrm{scatter\_add}(\bs{f}^{e}_{\mathrm{int}},\mathcal{E})$ \Comment{$(n_{\mathrm{nod}},3)$, reactions at constrained DOFs}
\State $\bs{R} \leftarrow -(1-\Gamma_D)\odot\bs{f}_{\mathrm{int}}$ \Comment{residual at free DOFs ($\bs{f}_{\mathrm{ext}}=\bs{0}$)}
\State $\mathcal{L}_{R} \leftarrow \sum_{i,k} (R_{ik}/s)^2$

\Statex \textbf{Equilibrium check} (no gradient)
\State $R_{\max} \leftarrow \max_{i,k}|R_{ik}|$
\State $\mathcal{A} \leftarrow \{(i,k):\ |f_{\mathrm{int},ik}|>\epsilon\max_{j,l}|f_{\mathrm{int},jl}|\}$ \Comment{active components, constrained included}
\State $\tilde{q} \leftarrow \frac{1}{2}\Big(\tilde{q}_{\mathrm{stored}} + \frac{1}{|\mathcal{A}|}\sum_{(i,k)\in\mathcal{A}}|f_{\mathrm{int},ik}|\Big)$
\State $r_{\max} \leftarrow R_{\max}/\tilde{q}$

\State \textbf{return} $\mathcal{L}_{R}$, $\mathcal{L}_{\mathrm{bc}}$, $\mathcal{L}_{\mathrm{neg}J}$, $r_{\max}$
\end{algorithmic}
\end{algorithm}
\end{document}